\documentstyle[aps,preprint,epsfig]{revtex}

\def \beq {\begin{equation}}
\def \eeq {\end{equation}}

\begin{document}

\draft

\title{Chaos around the superposition of a monopole and a thick disk}

\author{Alberto Saa\footnote{e-mail: {\tt asaa@ffn.ub.es}. Permanent
address: Departamento de Matem\'atica Aplicada, \\ IMECC -- UNICAMP,
        C.P. 6065, 13081-970 Campinas, SP, Brazil.}}
\address{
Departament de F\'\i sica Fonamental,  Universitat de Barcelona, \\
Av. Diagonal, 647, 
08028 Barcelona, Spain.}

\maketitle

\begin{abstract}
We extend recent investigations on the integrability
of oblique orbits of test particles under the gravitational
field corresponding to the superposition of an
infinitesimally thin disk and a monopole to the
more realistic case, for astrophysical purposes, of a thick disk. 
Exhaustive numerical analyses were performed and 
the robustness of the
recent results is confirmed.
We also found 
that, for smooth distributions
of matter, the disk thickness can attenuate the chaotic behavior
of the bounded oblique orbits. 
Perturbations leading to the breakdown of the reflection
symmetry about the equatorial plane, nevertheless,
may enhance significantly the chaotic behavior,
in agreement with recent studies on oblate models.

\end{abstract}

The recent  observational evidences suggesting 
that huge black-holes might inhabit the center of many active 
galaxies\cite{kor} have motivated some 
 investigations on the dynamics of test particles in gravitational
systems consisting in the superposition of monopoles and disks.
Infinitesimally thin disks are frequently used to model
flattened galaxies\cite{saslaw}.  Some exact relativistic
solutions describing the superposition of non-rotating black-holes
and static thin disk, and their respective Newtonian limits,
were presented and discussed in \cite{lele}.
In \cite{saave}, the integrability of oblique orbits of test
particles around the exact superposition of a black-hole
and an infinitesimally static
 thin disk was considered. Bounded zones of chaotic behavior
were found for both the relativistic and Newtonian limits.
There are several examples in the literature of chaotic motion
involving black-holes: in the fixed two centers problem\cite{2b,2b1,2b2},
in a black-hole surrounded by gravitational waves\cite{bw,bw1}, 
and in several core--shell models with
relevance to the description of galaxies (see \cite{vl} for a recent
review). As to the Newtonian case, we notice, for instance,
 the recent work
of C. Chicone, B. Mashhoon, and D. G. Retzloff on the chaotic
behavior of the Hill system\cite{hill}.

The Newtonian analysis of \cite{saave} has revealed an
interesting property of the dynamics of oblique bounded
orbits around the superposition of a monopole and an infinitesimally thin
disk. Since one was manly interested in bounded
motions close to the monopole, it was assumed that the disk was infinite
and homogeneous. This situation corresponds the simplest  superposition
of a monopole and a disk. Using cylindrical coordinates 
$(r,\theta,z)$ with the monopole,
with mass $M$,
located in the origin and the disk corresponding to the plane $z=0$, the
gravitational potential is given by 
\beq
\label{pot}
V(r,\theta,z) = -\frac{M}{\sqrt{r^2 + z^2}} + \alpha |z|,
\eeq
where $\alpha$ is a positive parameter standing  for 
the superficial mass density of the disk. The
angular momentum $L$ in the $z$ direction is conserved, and we can easily
reduce the three-dimensional original problem to a two-dimensional
one in the coordinates $(r,z)$ with the Hamiltonian given by
\beq
\label{ham}
H = \frac{\dot{r}^2}{2} + \frac{\dot{z}^2}{2} + \frac{L^2}{2r^2} 
 -\frac{M}{\sqrt{r^2 + z^2}} + \alpha |z|.
\eeq
The Hamiltonian (\ref{ham}) is smooth everywhere except on
the plane $z=0$.
Moreover, the parts of the trajectories restricted to the regions
$z>0$ and $z<0$ are integrable\cite{saave}. The corresponding
Hamilton-Jacobi equations restricted to these regions 
can be properly separated in parabolic coordinates, leading,
respectively, to
the second constants of motion
\beq
\label{C1}
C_{z>0} = 
R_z - \alpha\frac{r^2}{2}
\eeq
and
\beq
\label{C2}
C_{z<0} =
R_z + \alpha\frac{r^2}{2},
\eeq
where $R_z$ is the $z$ component of the Laplace-Runge-Lenz vector.
Note that $C$ is not smoothly defined on the disk.
With the two constant of motions $H$ and $C$, the equations for the
trajectories of test particles, restricted to the
regions $z>0$ and $z<0$, can be properly reduced to quadratures in
parabolic coordinates\cite{dor,dor1}. A complete bounded oblique trajectory,
therefore, corresponds to the matching of an infinite number
of integrable trajectory pieces. Hence, the widespread zones of chaotic
motion detected in \cite{saave} have their origin in the changes in the
value of the constant $C$  when the test particle crosses the
disk $z=0$.  
As to the relativistic case, in contrast, 
the  trajectory pieces 
that do not cross the disk are themselves
non-integrable, leading to typically larger
chaotic regions than in the corresponding Newtonian limit\cite{saave}. 

Here, we study the robustness of the results for the
Newtonian case by considering 
the more realistic case, for the description of
flattened galaxies, of a superposition of a central
monopole and  a 
smooth thick disk with the potential (\ref{pot}) as the vanishing
disk thickness limit. A smooth 
distribution of matter is considered  as a disk
if its radial gradients are much smaller than its vertical
ones. A minimally realistic model for a rotating thick disk should 
obey Emden's equation for the stability of rotating
polytropes\cite{saslaw}. As in \cite{saave}, we will neglect
the radial gradients, and, in this case, Emden's equation for the
disk matter density $\rho(z)$ states that $(G=1)$
\beq
\label{emde}
\kappa\lambda\rho^{\lambda-2}\rho'' 
+\kappa\lambda(\lambda-2)\rho^{\lambda-3} (\rho')^2 = - 4\pi \rho,
\eeq
where $\kappa$ is the parameter relating the pressure to the matter density 
in the polytropic equation of state and 
$\lambda=1+1/n$, $n$ being the polytrope index. 
The matter density $\rho$ is assumed to obey Poisson's equation
$\nabla^2 V_{\rm D} = 4\pi  \rho$. For the isothermal case 
($\lambda = 1$), equation (\ref{emde}) admits as
a solution the following distribution of matter
\beq
\label{rho}
\rho(z) = \frac{\alpha}{4\pi  z_0}{\rm sech}^2\frac{z}{z_0},
\eeq
which corresponds to the potential
\beq
\label{pot1}
V_{\rm D}(z) = \alpha z_0 \ln \cosh\frac{z}{z_0},
\eeq
where $z_0$ measures the disk ``thickness'', and
$\alpha$ its ``superficial'' mass density. They obey the
relation $2\kappa = \alpha z_0$. Typically, for realistic models of
a rotating dust disk one has 
$z_0^2 \propto \left< V_z^2\right>$\cite{spitzer}. The matter
density (\ref{rho}) corresponds, therefore,
to a stable and smooth distribution of rotating matter concentrated on
the plane $z=0$. Moreover, in the limit of $z_0\rightarrow 0$,
we recover from (\ref{pot1}) the infinitesimal thin disk potential 
$V_{\rm D}(z) = \alpha |z| $ and the
corresponding $\delta$ distribution of matter from (\ref{rho}).  

Thus, the dynamics of test particle moving around the superposition
of our smooth thick disk and a monopole will be governed by the
following smooth Hamiltonian
\beq
\label{ham1}
H = \frac{\dot{r}^2}{2} + \frac{\dot{z}^2}{2}+ \frac{L^2}{2r^2} 
 -\frac{M}{\sqrt{r^2 + z^2}}  + \alpha z_0
\ln\cosh \frac{z}{z_0}.
\eeq
We wish to stress that the potential in (\ref{ham1}) corresponds,
indeed, to a first approximation of a realistic superposition
of a monopole and a smooth thick disk with matter distribution
given by (\ref{rho}). The whole superposition must also obey Emden's
equation; in the present case it fails to obey at the origin. We
are neglecting the radial gradients of the matter 
distributions caused by the stresses induced
by the central monopole.

The Hamiltonian (\ref{ham1}) does not belong to the class of  
integrable two-dimensional potentials with a second constant of motion
polynomial in the momenta\cite{prc}. We will
present strong evidences that (\ref{ham1}) does not have a
second constant of motion at all.
We could solve numerically the system governed by (\ref{ham1}) with
great accuracy. Figure 1 shows a typical 
Poincar\'e's section
$(H=-0.2, L=M=1, \alpha = 0.1, z_0 = 1.5)$  across the plane $z=0$ 
revealing a widespread chaotic behavior. The disk thickness
$z_0$, in this case, has the same  magnitude of the typical
$z$-amplitude of the trajectories.
Figure 2 shows a sequence of low-energy 
sections
$(H=-0.3, L=M=1, \alpha = 0.1, z_0 = 0.0 \ {\rm (a)}, 
                                     0.1 \ {\rm (b)},  
                                     0.25 \ {\rm (c)}, \ {\rm and} \  
                                     0.5 \ {\rm (d)} )$, 
constructed
from the same trajectory initial conditions,
where the
attenuation of the chaotic behavior due to the disk thickness
can be clearly appreciated.
We could obtain thousands of intersections for each trajectory
with a cumulative error, measured by the constant $H$,
inferior to $10^{-12}$.
We notice that the equation of motion are invariant under the
following rescalings:
\begin{eqnarray}
r \rightarrow \lambda r, \quad &
z \rightarrow \lambda z, \quad &
t \rightarrow \lambda^{3/2} t, \nonumber \\
\alpha \rightarrow \lambda^{-2}\alpha, \quad &
M \rightarrow M, \quad &
z_0 \rightarrow \lambda z_0 \nonumber \\
H\rightarrow\lambda^{-1}H, \quad &
L \rightarrow\lambda^{1/2} L;  \quad&
\ \  
 \end{eqnarray}
and
\begin{eqnarray}
r \rightarrow \lambda' r, \quad &
z \rightarrow \lambda' z, \quad &
t \rightarrow \lambda' t, \nonumber \\
\alpha \rightarrow \lambda'^{-1}\alpha, \quad &
M \rightarrow \lambda' M, \quad &
z_0 \rightarrow \lambda' z_0 \nonumber \\
H\rightarrow H, \quad &
L \rightarrow \lambda' L;  \quad&
\ \  
\end{eqnarray}
$\lambda >0$ and $\lambda' >0$, leading that, for each
triple of nonzero parameters $(M,\alpha,z_0)$, one has, in fact,
only one free parameter, namely $z_0\sqrt{\alpha/M}$.

The limit of $z_0$ much larger than the typical $z$-amplitude
of the trajectories deserves a special attention. For this case,
the Hamiltonian can be well approximated by
\beq
\label{ham2}
H_\infty = \frac{\dot{r}^2}{2} + \frac{\dot{z}^2}{2}+ \frac{L^2}{2r^2} 
 -\frac{M}{\sqrt{r^2 + z^2}}  + \frac{\beta}{2} z^2,
\eeq
where $\beta = \alpha/z_0$. Such a potential is related to
the potential of the Kepler problem perturbed by a 
quadrupole halo potential considered in \cite{vl}. Figure 3
shows a typical Poincar\'e's section 
$(H_\infty =-0.15, L=M=1, \beta = 0.1)$
across the plane $z=0$ 
revealing a widespread chaotic behavior for the system
governed by (\ref{ham2}). 
We could also obtain thousands of intersections for each trajectory
with a cumulative error inferior to $10^{-12}$.
Like in the infinitesimally thin disk case\cite{saave}, 
due to the existence of two rescaling invariances, this Poincar\'e's section 
can be obtained for any non-zero values of $\beta$ and $M$. 

Our numerical finds confirm the robustness of the
results presented in \cite{saave}. The chaotic behavior of
bounded orbits can be considered as inherent to any system
consisting in the superposition of a disk and a central monopole.
We stress that the superpositions we have considered are
symmetric under the reflection about the equatorial plane. A
perturbation leading to the breakdown of the reflection
symmetry could increase significantly the chaotic behavior,
as the recent studies on oblate models  have suggested\cite{vl}.
We could indeed check such fact by considering a
dipole-like perturbation of the Hamiltonian (\ref{ham1})
\beq
\tilde{H} =  H - D\frac{z}{(r^2 + z^2)^{3/2}}.
\eeq
Figure 4 presents a typical Poincar\'e's surface 
$(\tilde{H}=-0.3, L=M=1, \alpha = 0.1, z_0 = 0.25, D = 5\times 10^{-2})$ 
for this case. The zones of chaotic motion are larger
than in the corresponding case for which $D=0$
(Fig 2.c). However, the central family of regular orbits,
if it does exit for $D=0$, seems to be robust against
dipole-like perturbations. Analogous conclusions hold also
for the large $z_0$ case.

\acknowledgements

The author is grateful to CNPq and FAPESP for the financial support, and to
Prof. P.S. Letelier and R. Venegeroles
for stimulating discussions.

\begin{figure}
\vspace{0.5cm}
\caption{Poincar\'e's section $(r,\dot{r})$ across the plane $z=0$ for 
oblique orbits, with $H=-0.2$ and $L=1$, 
around the superposition of a monopole with
mass $M=1$ and a smooth disk with surface
mass density $\alpha=0.1$ and thickness $z_0=1.5$. }
\end{figure}

\newpage

\begin{figure}
\epsfxsize=7.5cm
\epsfxsize=7.5cm

\epsfxsize=7.5cm
\epsfxsize=7.5cm

\vspace{0.5cm}
\caption{Sequence of Poincar\'e's section $(r,\dot{r})$ 
across the plane $z=0$ for 
oblique orbits,
with $H=-0.3$ and $L=1$,
 around the superposition of a monopole with
mass $M=1$ and a smooth disk with mass surface
density $\alpha=0.1$ and thickness $z_0 = 0.0$ (a),
$z_0 = 0.1$ (b), $z_0=0.25$ (c), and $z_0 = 0.5$ (d). The
sections were constructed from the same trajectory initial conditions.}

\end{figure}

\begin{figure}
\vspace{0.5cm}
\caption{Poincar\'e's section $(r,\dot{r})$ across the plane $z=0$ for 
oblique orbits, with $H_\infty=-0.15$ and $L=1$, 
around the superposition of a monopole with
mass $M=1$ and a smooth disk of large thickness with 
$\beta=0.1$.}
\end{figure}

\newpage

\begin{figure}
\vspace{0.5cm}
\caption{Poincar\'e's section $(r,\dot{r})$ across the plane $z=0$ for 
oblique orbits, with $\tilde{H}=-0.3$ and $L=1$, 
around the superposition of a central source characterized by a 
monopole charge
$M=1$ and a dipole strength $D=5\times 10^{-2}$,
 and 
a smooth disk with mass surface
density $\alpha=0.1$ and thickness $z_0 = 0.25$. The case
corresponding to $D=0$ is presented in Fig. 2.c.}
\end{figure}

\end{document}